\documentclass[journal]{IEEEtran}

\usepackage{amssymb}
\usepackage{amsmath}
\usepackage{graphicx}
\newcommand{\dv}{d_\mathrm{v}}
\newcommand{\dc}{d_\mathrm{c}}
\newcommand{\SN}{\mathtt{S}}
\newcommand{\SR}{\mathtt{SR}}
\newcommand{\SD}{\mathtt{SD}}
\newcommand{\RD}{\mathtt{RD}}
\newcommand{\RN}{\mathtt{R}}
\newcommand{\DN}{\mathtt{D}}
\newcommand{\GF}{\mathrm{GF}}
\newcommand{\BC}{\mathrm{BC}}
\newcommand{\MAC}{\mathrm{MAC}}

\usepackage{color}	% required for `\textcolor' (yatex added)
\begin{document}

%
% paper title
% can use linebreaks \\ within to get better formatting as desired
\title{Simple Low-Rate Non-Binary LDPC Coding for Relay Channels}
%
%

%
%
% author names and IEEE memberships
% note positions of commas and nonbreaking spaces ( ~ ) LaTeX will not break
% a structure at a ~ so this keeps an author's name from being broken across
% two lines.
% use \thanks{} to gain access to the first footnote area
% a separate \thanks must be used for each paragraph as LaTeX2e's \thanks
% was not built to handle multiple paragraphs
%

\author{Puripong~Suthisopapan,~\IEEEmembership{Student Member,~IEEE,}
        Kenta~Kasai,~\IEEEmembership{Member,~IEEE,} \\
        Anupap Meesomboon, 
        Virasit Imtawil, 
        and~Kohichi Sakaniwa,~\IEEEmembership{Senior Member,~IEEE,}
        }

%% The paper headers
%\markboth{Journal of \LaTeX\ Class Files,~Vol.~6, No.~1, January~2007}%
%{Shell \MakeLowercase{\textit{et al.}}: Bare Demo of IEEEtran.cls for Journals}

\maketitle

%\begin{abstract}
%\textcolor{magenta}{A} Binary LDPC coded relay system has been studied previously with the assumption of infinite codeword length.
%In this paper, we deal with non-binary LDPC codes which can outperform their binary counterpart especially for practical codeword length.
%We utilize non-binary LDPC codes and recently invented non-binary coding techniques 
%\textcolor{magenta}{known as multiplicative repetition and puncturing with recoverable step}
%to design the simple coding strategies for \textcolor{magenta}{the} decode-and-forward half-duplex relay channel.
%We claim that our proposed strategies are simple 
%since \textcolor{magenta}{the} destination and \textcolor{magenta}{the} relay can decode at almost the same complexity 
%by sharing the same structure of decoder.
%Numerical experiments are carried out to demonstrate that 
%the proposed strategies offer better error rate than the previously found scheme even with the shorter codeword length.
%Comparing to the theoretical limits, the performances obtained by non-binary LDPC coded relay systems surpass the capacity of direct transmission and 
%NBLDPC codes also work within less than $1.5$dB from the achievable rate of \textcolor{magenta}{the} relay channels.
%\end{abstract}

\begin{abstract}
Binary LDPC coded relay systems have been well studied previously with the assumption of infinite codeword length.
In this paper, we deal with non-binary LDPC codes which can outperform their binary counterpart especially for practical codeword length.
We utilize non-binary LDPC codes and recently invented non-binary coding techniques known as \textit{multiplicative repetition} 
to design the low-rate coding strategy for the decode-and-forward half-duplex relay channel.
We claim that the proposed strategy is simple 
since the destination and the relay can decode with almost the same computational complexity 
by sharing the same structure of decoder.
Numerical experiments are carried out to show that the performances obtained by non-binary LDPC coded relay systems surpass the capacity of direct transmission 
and also approach within less than $1.5$ dB from the achievable rate of the relay channels.
\end{abstract}

% Note that keywords are not normally used for peerreview papers.

% IEEEtran.cls defaults to using nonbold math in the Abstract.
% This preserves the distinction between vectors and scalars. However,
% if the conference you are submitting to favors bold math in the abstract,
% then you can use LaTeX's standard command \boldmath at the very start
% of the abstract to achieve this. Many IEEE journals/conferences frown on
% math in the abstract anyway.

% no keywords

% For peer review papers, you can put extra information on the cover
% page as needed:
% \ifCLASSOPTIONpeerreview
% \begin{center} \bfseries EDICS Category: 3-BBND \end{center}
% \fi
%
% For peerreview papers, this IEEEtran command inserts a page break and
% creates the second title. It will be ignored for other modes.
%\IEEEpeerreviewmaketitle

\section{Introduction}
%\subsection{We focus in good things}
Over recent years, cooperative communication has been studied extensively in order to increase the capacity of wireless  networks \cite{coop1,coop2,coop3}.
%The terminals in cooperative network work together to send their information.
In cooperative networks, terminals act as relays which can forward the received information to destinations or terminals.
It is well known that spatial diversity is the promising technique for improving the quality and reliability of wireless links \cite{diversity}.
The spatial diversity can be obtained by adopting multiple antennas in a transmitter or/and a receiver \cite{mimo1,mimo2}.
Relays in a cooperative network form a virtual multiple antenna system and hence allow a single-antenna user to transmit the information through the different and independent paths.
Without employing multiple antennas, the spatial diversity still can be achieved via relaying.
Therefore, the concept of relaying is applicable to mobile terminals which cannot support multiple antenna.
However, cooperative communications with multiple antennas also have been studied to achieve higher capacity \cite{mimo_coop1,mimo_coop2} but this is beyond the scope of this paper.

A single relay channel is an elementary component of a cooperative network \cite{relay_chan1}.
The single relay channel simply consists of three terminals which are a source, a relay and a destination.
It is theoretically shown that the achievable rate of a single relay channel is much higher than the capacity of direct transmission \cite{relay_chan2}.
Many protocols, such as amplify-and-forward \cite{relay_chan3} or compress-and-forward \cite{relay_chan4}, have been developed for processing the signals in relay channels.
Previous works have shown that the decode-and-forward protocol achieves better performance compared with other protocols when a quality of source-relay link is good \cite{relay_chan5,relay_chan6}.
By using the decode-and-forward protocol, a relay is capable to decode the received signal 
and forward the decoded information to the destination.

At the present moment, half-duplex relaying is regarded more practical
since full-duplex relaying cannot be implemented efficiently.
The operation in a half-duplex scheme is divided into two modes named as broadcast (BC) and multiple access (MAC) modes.
In BC mode, the source transmits signal to both the relay and the destination.
The relay does not have its own information to transmit. 
Thus the relay only listens to the source in BC mode.
In MAC mode, both the relay and the source transmit their signals to the destination.
In this paper, we focus only on the half-duplex relay channel with the decode-and-forward protocol 
and restrict ourselves to the case of a single relay.

%\subsection{Related works}
Several binary LDPC (BLDPC) coding strategies have been developed to enhance the performance of a decode-and-forward half-duplex relay channel \cite{code_relay1,code_relay2}.
The main challenging problem in this field is that one needs to design a coding strategy that works at both source-relay and source-destination links.
In other words, a code-structure is necessary to exhibit excellent performance at two different signal to noise ratios.
After designing the code properly, the code is distributed to the source, the relay and the destination.
%In literatures, we known that low-density parity-check (LDPC) codes are the capacity-approaching channel code for AWGN channel \cite{LDPC}. 
%Therefore, code designers are interested to develop the distributed LDPC codes for a relay channel. 
We briefly introduce some prominent works in this research field.
Chakrabarti \textit{et al}. optimized the degree profile of BLDPC codes for half-duplex relay channel by mean of density evolution \cite{LDPC_relay1}.
Razaghi and Yu developed a technique called bi-layer BLDPC codes \cite{LDPC_relay2} in which overall graph of an BLDPC code is subdivided into lower and upper parts corresponding to codes for source-relay and source-destination links.
In \cite{code_relay2}, Li \textit{et al}. proposed the distributed BLDPC code using a rate-compatible structure and also developed a method which can accurately predict the performance of BLDPC codes for a relay channel.
Recently, Cances \textit{et al}. extended the idea of bi-layer and designed the BLDPC codes that work closely to theoretical limit \cite{LDPC_relay3}.
We note that all works mentioned above focused on BLDPC codes.

%\subsection{Describe The Problem}
Although the significant progress has been made to design the distributed LDPC codes for the half-duplex relay channel. However, there still exist some problems that can be addressed as follows: 
1) The design of optimized LDPC codes for the relay channel requires a task to obtain such an optimized degree profile. 
2) The sophisticated degree profile of LDPC code obtained from an optimization process makes the hardware implementation quite complicated.
3) LDPC codes with the optimized degree profile require very large codeword length (e.g. $10^5$ bits) to perform close to the theoretical limit.
The large codeword length leads to the transmission latency and complex decoders. 
which are not preferred in the real world communication.
Moreover, the good performance of optimized LDPC code are not guaranteed when the codeword length is short or moderate.
In this paper, we overcome these problems by considering non-binary LDPC (NBLDPC) codes.

%\subsection{Our Tool}
An NBLDPC code is defined by a sparse parity-check matrix defined over $\GF(2^m)$, $m>1$ \cite{nb1}.
Over point-to-point channels, it was shown in \cite{nb1,nb2} that the NBLDPC codes can outperform their binary counterparts especially for the short and moderate codeword lengths.
Therefore, the NBLDPC codes can be employed to improve the performance of practical wireless communications since the short and moderate codeword lengths (e.g. a few thousand bits) are now currently used \cite{std1,std2}.
Good NBLDPC codes defined over higher order Galois fields tend to have the regular degree profile (row and column weights of parity-check matrix are constant) \cite{nb4}.
The regular NBLDPC codes defined over $\GF(2^m)$ with column weight 2 are empirically known \cite{nb4} as the best performing codes for $2^m>64$.
Therefore, the first and second problems addressed in the previous paragraph can be alleviated since we can use NBLDPC codes with regular degree profile for transmission and hence do not need to optimize the degree profile of NBLDPC codes.

%\subsection{What's new in this topic?}
At this moment, little progress has been made for an application of NBLDPC codes on relay channels but 
the aforementioned reasons mentioned earlier motivate us to design the distributed NBLDPC codes for relay channel.
%The main contribution of this paper is to propose the use of the NBLDPC codes for a relay channel.
%\subsection{Deliver Our Idea to Reader}
This paper, we develop the low-rate NBLDPC coding strategy for the decode-and-forward half-duplex relay channel.
We apply the concept of \textit{multiplicative repetition} \cite{MR1,MR2} to design the distributed low-rate NBLDPC codes 
for the relay channel.
%We focus our attention on designing low-rate NBLPDC codes for the relay channel since the acheivable.
%Furthermore, we also apply rate-compatible NBLDPC codes punctured by the concept of \textit{recoverable step} \cite{punc1,punc2} to design the distributed high rate NBLDPC codes for the relay channel.
%\subsection{Our Contributions}
Based on the proposed NBLDPC coding strategies, 
the relay and the destination can perform the decoding process with the same computational complexity 
even the coding rate of the destination is much lower than that of the relay.
% main results
Finally, we show that the proposed strategy provides the relay systems with very good decoding performances 
especially for the short and moderate codeword lengths.

%\subsection{Paper Organization}
The rest of this paper is organized as follows. 
Section II, we explain the half-duplex relay channel in time-division mode together with the decode-and-forward protocol. 
The NBLDPC codes are introduced in Section III. 
Section IV, we develop the simple NBLDPC coding strategy which applicable to low-rate relay channel.
In Section V, we present the decoding performance of the proposed coding strategy.
The conclusions are finally given in Section VI.

\section{Introduction to NBLDPC codes}
Before describing the proposed coding strategy, we first introduce NBLDPC codes, multiplicative repetition, puncturing
and  their decoding algorithm. 
\subsection{NBLDPC codes}
NBLDPC codes are defined by sparse parity-check matrices defined over $\GF(2^m)$, $m>1$ \cite{nb1}.
We can represent each element $x \in \GF(2^m)$ by using binary sequences of length $m$ bits.
For transmitting over the binary input channels, each non-binary symbol in $\GF(2^m)$ needs to be represented by a binary sequence of length $m$.
For each $m$, we fix a Galois field $\GF(2^m)$ with a primitive element $\alpha$ and its primitive polynomial $\pi$. 
Once a primitive element $\alpha$ of $\GF(2^m)$ is fixed, each symbol is given a $m$-bit representation \cite[pp.~110]{macwilliams77}.
For example, with a primitive element $\alpha\in\GF(2^3)$ such that $\pi(\alpha)=\alpha^3+\alpha+1=0$, each symbol is represented as
$0=(0,0,0)$, $1=(1,0,0)$, $\alpha=(0,1,0)$, $\alpha^2=(0,0,1)$,
$\alpha^3=(1,1,0)$, $\alpha^4=(0,1,1)$, $\alpha^5=(1,1,1)$ and $\alpha^6=(1,0,1)$.
Assigning of binary sequence to each element depends on the primitive polynomial used to construct the field.
Specifically, an NBLDPC code $C$ over $\GF(2^m)$ is defined by the null-space of a sparse $M \times N$ parity-check matrix $\mathbf{H}={(h_{i_{j}})}$ defined over $\GF(2^m)$
\begin{equation}
\label{eq6}
C = \{ \mathbf{x} \in \GF(2^m)^{N} \mid \mathbf{H}\mathbf{x} = \mathbf{0} \in \GF(2^m)^{M}\},
\end{equation}
where $\mathbf{x} = (x_1,\ldots,x_N)$ is the codeword.
The $c$-th parity-check equation for $c=1,\ldots,M$ is written as
\begin{equation}
\label{eq7}
h_{c_{1}}x_1 + h_{c_{2}}x_2 + \cdots h_{c_{N}}x_{N} = 0,
\end{equation}
where $h_{c_{1}},\ldots,h_{c_{N}} \in \GF(2^m)$ are the entries of $c$-th row of $\mathbf{H}$.
The parameter $N$ is the codeword length in symbol.
Throughout this paper (assuming $\mathbf{H}$ is of full rank), 
we denote the number of information symbols by $K=N-M$.
Let $n = mN$ and $k = mK$ be the codeword length and information length in bit, respectively.
The coding rate $R$ of any NBLDPC code can be computed by $R = 1 - \dv/\dc = K/N$.

\begin{figure}[htb]
\centering
\includegraphics[scale=0.45]{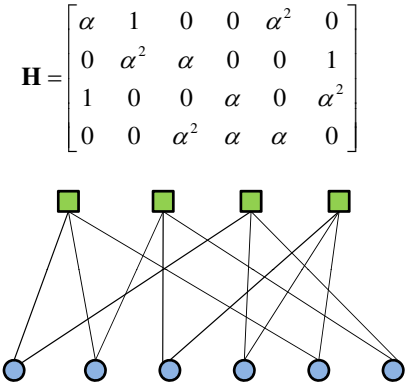}
\caption{An example of $(2,3)$-regular NBLDPC codes. The codeword length of this code is $N=6$. The number of parity symbol is $M=4$. The upper part shows the parity-check matrix of size $4\times6$. The lower part shows the corresponding Tanner graph of this parity-check matrix but here we omitted edge labels for simplicity. Circle and square nodes represent variable and check nodes, respectively.}
\label{nb_ldpc}
\end{figure}

An NBLDPC code is called $(\dv, \dc)$-regular if the parity-check matrix of the code has constant column weight $\dv$ and row weight $\dc$.
In this paper, only $(\dv=2,\dc)$-regular NBLDPC codes defined over $\GF(2^8)$ are considered, 
since they are empirically known as the best performing codes. 
The $\dv, \dc$ indicates the column and row weights of parity-check matrix, respectively.
We can represent the parity-check matrix of an NBLDPC code by a \textit{Tanner graph} with variable and check nodes \cite{mct_book}.
Each variable node and check node represents a coded symbol and a parity-check equation, respectively.
Therefore, the number of variable nodes and check nodes are equal to $M$ and $N$, respectively.
Figure \ref{nb_ldpc} shows an example of the parity-check matrix of a $(2,3)$-regular NBLDPC code defined over $\GF(2^2)$ and the corresponding Tanner graph.

Since, in this paper, we focus only on the binary transmission,
we describe how to transmit the NBLDPC coded symbols $x_v$ by using the binary modulation scheme 
for $v=1,2,\ldots,N$ and $i=1,2,\ldots,m$.
At the $v$-th output of NBLDPC encoder, 
each coded symbol $x_{v} \in \GF(2^m)$ is mapped to 
the binary sequence of $m$ bits $(x_{v,1},x_{v,2},\ldots,x_{v,m}) \in \GF(2)^m $
according to the primitive polynomial as described above.
The obtained binary sequence $(x_{v,1},x_{v,2},\ldots,x_{v,m}) \in \GF(2)^m$ is then mapped to $m$ modulated signals
through the mapper for the transmission.

\subsection{Multiplicative Repetition}
More recently, Kasai \textit{et al}. proposed an efficient method called $\textit{multiplicative repetition}$ to construct low-rate NBLDPC codes \cite{MR1,MR2}.
Over point to point channels, 
the low-rate codes constructed by this method outperform the previously found low-rate codes.
By using the NBLDPC code of rate $R_1$, 
we can easily construct code NBLDPC codes of lower rate $R<R_1$ as described as follows.

Let $C_1$ denotes a NBLDPC code of length $N$ and coding rate $R_1$.
Since code $C_1$ is used to constructed another codes, we refer to $C_1$ as a \textit{mother code}.
A low-rate code $C_2$ of length $2N$ and coding rate $R_2 = \frac{1}{2}R_1$ can be constructed as follows.
We select $N$ coefficients $r_{N+1},\ldots,r_{2N}$ randomly from $\GF(2^m)\setminus\{0\}$.
Note that we define $r_v=1\in\GF(2^m)$ for $v=1,\dotsc,N$ for simplicity of notation.
We then multiplicatively repeat the coded symbols from $C_1$ with the coefficients to obtain the lower-rate code $C_{2}$
as follows.
\begin{center}
$ C_2 = \{ (x_1,\ldots,x_{2N}) \in \GF(2^m)^{2N} \mid x_{N+v} = r_{N+v}x_{v} $, 

~~~~~for $v=1,\ldots,N$, $(x_1,\ldots,x_N) \in C_1$ \}.
\end{center}
In this way, we can construct a $(2N,K)$ code from an $(N,K)$ code.

The codes $C_3,C_4,\ldots,C_T$ of lower coding rates can also be constructed from code $C_1$ through the 
multiplicative repetition process.
We refer a parameter $T$ as \textit{repetition parameter}.
For $T\geq3$, in a recursive fashion, $N$ coefficients $r_{(T-1)N+1},\ldots,r_{TN}$ are chosen randomly from $\GF(2^m)\setminus\{0\}$.
The code $C_T$ of rate $R_T = \frac{1}{T}R_1$ can also be constructed as follows.
\begin{align*}
  C_T = \{ &(x_1,\ldots,x_{TN}) \in \GF(2^m)^{TN} \mid \\
& x_{(T-1)N+v} = r_{(T-1)N+v}x_{v},  \text{ for } v=1,\ldots,N, \\
& (x_1,\ldots,x_{(T-1)N}) \in C_{T-1} \}.
\end{align*}
Therefore, we can easily construct $(TN,K)$ code from $(N,K)$ code in a recursive fashion.

\begin{figure}[htb]
\centering
\includegraphics[scale=0.7]{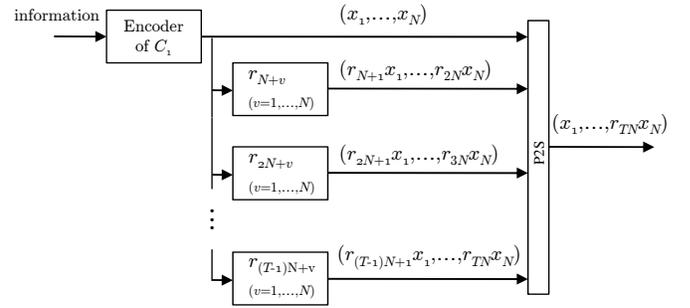} 
\caption{Block diagram of multiplicatively repeated encoder of $C_T$. In this figure, P2S represents parallel to serial operation.}
\label{mr_encoder}
\end{figure}

For example, we can construct $C_2,C_3,\ldots,C_T$ with coding rate $1/6,1/9,\ldots,1/3T$ respectively 
when the mother code $C_1$ is a $(2,3)$-regular NBLDPC code.
Figure \ref{mr_encoder} shows the block diagram of multiplicatively repeated encoder of $C_T$.
%The code structure obtained by this method can be visualized in Fig. \ref{mr_structure}. 
We will show in the next section that the multiplicative repetition 
is applicable to design low-rate coding strategy for relay channels.
%\begin{figure}[htb]
%\centering
%\includegraphics[scale=1.05]{fig6_mr_code_structure.eps} 
%\caption{The structure of a multiplicatively repeated code. $\mathrm{mr}_{i}(C_1)$ represents the multiplication between codeword form $C_1$ and randomly chosen multiplicative coefficients where $2\leq i \leq T$.}
%\label{mr_structure}
%\end{figure}

%In this paper, we select NBLPDC codes defined over $\GF(2^m)$ with $(\dv,\dc)$-regular structure as channel code for relay transmission.
%NBLPDC coding strategy for relay channel will be introduced in the next section.

\subsection{Puncturing}
By \textit{puncturing} a code, we obtain a higher-rate code.
We cancel some parity symbols produced from the encoder of mother code $C$ of rate $R =\frac{K}{N}$ 
to form the code $C'$ of higher rate $R' = \frac{K}{N-N_p}$ 
where $N_p$ is the number of punctured symbols.
We refer to the code $C$ as the \textit{mother} code.
The puncturing plays an important role in constructing the \textit{rate-compatible} codes.
The rate-compatible codes are very useful in wireless communications 
which employ hybrid automatic repeat request (HARQ) protocol.
The rate-compatible codes can flexibly adapt the coding rate according to the channel quality 
by adopting only one mother code.

\begin{figure}[htb]
\centering
\includegraphics[scale=0.5]{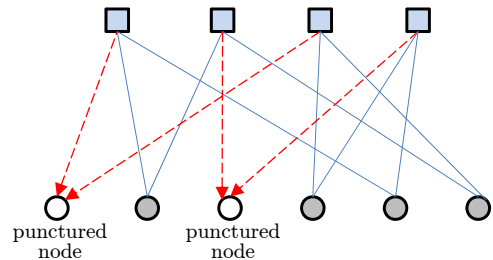}  
\caption{An example of puncturing on $(2,3)$-regular NBLDPC codes with one step recoverable. The codeword length of this code is $N=6$. The number of parity symbol is $M=4$. Circle and square nodes represent variable and check nodes respectively. We intend to puncture the variable node with one step recoverable first and then continue to puncture variable node with two step recoverable and so on.}
\label{ex_punc}
\end{figure}

Instead of random puncturing, we find the location of variable node for puncturing 
by the concept called \textit{recoverable step} \cite{punc1}.
This method helps the decoding process at $\RN$ because the punctured variable nodes will receive nonzero messages at small number of iterations.
We first search for the group of variable nodes with one-step-recoverable.
After that, we then search for the group of variable nodes with two-step-recoverable, and so on.
We therefore puncture variable nodes by the ascending order of recoverable steps.
An example of puncturing on the $(2,3)$-regular NBLDPC codes with one step recoverable 
is illustrated in Fig. \ref{ex_punc}.
In this paper, 
the concept of puncturing by recoverable step 
will be used to obtain the rate-compatible NBLDPC codes for the relay channel.

%For the NBLDPC codes, each NBLDPC symbol comprise of $m$ bits 
%and we have free choice to puncture NBLDPC symbol partially or entirely.
%Recently, Klinc \textit{et al}. proposed a method to puncture NBLDPC codes partially \cite{nb_punc_bit}.
%This work shows that the performance of punctured NBLDPC codes can be enhanced by using the concept of recoverable step together with puncturing partially.

\subsection{Decoding algorithm for NBLDPC codes}
With some abuse of notation, we denote the $v$-th variable node by $v$.
Let $X_v$ be the random variables with realization $x_v$  where $X_v$ represents the coded symbol and $v=1,2,\ldots,N$.
Let $Y_v$ be the random variables with realization $y_v$ 
which is the received value from the channel $\Pr(Y_{v}|X_{v})$.
The probability of transmitted symbol $\Pr(X_v)$ is assumed to be uniform.
We assume that the decoders at both relay and destination knows the channel transition probability
\begin{equation}
\label{tran_prob}
\Pr(X_v=x|Y_v=y_v), v=1,2,\ldots,N,
\end{equation}
for $x \in \GF(2^m)$. 
%$\mathbf{b}_{v} = (b_{((v-1)m+1)},\ldots,b_{((v-1)m+i)})$ 
For memoryless binary-input output-symmetric (MBIOS) channels, the channel transition probability is given as
\begin{equation}
\label{tran_prob_binary}
\Pr(X_{v}=x|Y_{v}=y_v) = \prod\limits_{i=1}^{m} \Pr(X_{v,i}=x_{i}|Y_{v,i}=y_{v,i}),
\end{equation}
where $X_{v,i}$ is the random variable of the transmitted bit, 
$(x_1,x_2,\ldots,x_m) \in \GF(2)^m$ is the binary representation of $x \in \GF(2^m)$ 
and the corresponding channel output $y_{v,i}$ and its random variable $Y_{v,i}$. 

%For $v=1,2,\ldots,N$, $a \in \{R,D\}$ and $b \in \{1,2\}$,
%We define the probability $p_v(x,y_{a,b})$ as follows
%\begin{equation}
%\label{def_func}
%p_v(x)  =  \xi \prod\limits_{i=1}^{m} \Pr(X_{v,i}=x_i|Y_{v,i}=y_{a,b}(m(v-1)+i))
%\end{equation}
%where $\xi$ is the normalization factor so that $\sum_{x \in \GF(2^m) } p_v(x,y_{a,b}) = 1$ 
%and $y_{a,b}(j)$ is the $i$-th component of received signal $y_{a,b}$.
%For example, $y_{\DN,1}(1)$ is the first component of received signal at destination in BC mode.
%This probability will be used as the input of decoder when the transmission takes place over MBIOS channels, 

The decoding of NBLDPC codes is accomplished on the Tanner graph through the belief propagation (BP) algorithm \cite{nb1,nb2}.
The BP algorithm exchanges probability vectors of length $2^m$ between variable nodes and check nodes at each iteration round $\ell$.
The BP algorithm mainly consists of 4 parts described as follows.

\textbf{initialization} : We set the iteration round $\ell = 0$.
Each variable node sends the initial message $p_{vc}^{(\ell=0)} = p_{v}^{(0)} \in \mathbb{R}^{2^m}$, 
to each adjacent check node $c$ where $v=1,2,\ldots,N$ and $c=1,2,\ldots,M$.
For general NBLDPC codes $(T=1)$ and multiplicatively repeated NBLDPC codes $(T\geq2)$, 
the probability $p_{v}^{(0)}(x)$ is initialized as follows
\begin{align*}
\label{init_NBLDPC}
  p_{v}^{(0)}(x)  =&     \Pr(X_{v}=r_{v}x|Y_{v}=y_{v})\\
 &  \xi\prod_{t=1}^{T}\Pr(X_{(tN+v}=r_{tN+v}x|Y_{tN+v}=y_{tN+v}),
\end{align*}
where $\xi$ represents the normalized constant so that 
$\sum_{x \in \GF(2^m) } p_{v}^{(0)}(x) = 1$.
Note that we defined $r_v=1\in\GF(2^m)$ for $v=1,\dotsc,N.$

Let $V_p$ denotes the set of punctured variable nodes.
For punctured NBLDPC codes and $v=1,2,\ldots,N$, the probability $p_{v}^{(0)}(x)$ is initialized as follows
\begin{equation}
\label{init_punc_NBLDPC}
p_v^{(0)}(x)  = \left\{ {\begin{array}{ll}
   { \frac{1}{{2^m }}} & {v \in V_p }  \\
   { \Pr (X_v  = x\left| {Y_v  = y_v } \right.)} & \textrm{otherwise}.
\end{array}} \right.
\end{equation}

\noindent\textbf{check to variable} : For each check node $c = 1,2,\ldots,M$, let $\partial_c$ be the set of adjacent variable nodes of $c$.
The check node $c$ sends the following message $p_{cv}^{(\ell)} \in \mathbb{R}^{2^m} $ to each adjacent variable node $v \in 
\partial_c$
\begin{align*}
&  \tilde{p}_{vc}^{(\ell)}(x) = {p}_{vc}^{(\ell)}(h_{vc}^{-1}x) \text{ for }x \in \GF(2^m),\\
& \tilde{p}_{cv}^{(\ell+1)} = \otimes_{v' \in \partial_c \setminus \{v\}  } \tilde{p}_{v'c}^{(\ell)},\\
& p_{cv}^{(\ell+1)}(x) = \tilde{p}_{cv}^{(\ell+1)}(h_{cv}x) \text{ for }x \in \GF(2^m),
\end{align*}
where $p_1 \otimes p_2 \in \mathbb{R}^{2^m} $ is a convolution of $p_1 \in \mathbb{R}^{2^m}$ 
which can be expressed as follows
\begin{align*}
 ( p_1 \otimes p_2 ) (x)  = \sum\limits_{\scriptstyle y,z \in \GF(2^m ) \hfill \atop \scriptstyle ~~~x = y + z \hfill} {p_1(y)p_2(z)} \text{ for }x \in \GF(2^m).
\end{align*}
The convolution appeared above can be efficiently calculated via FFT and IFFT \cite{nb2}.
Increment the iteration round as $\ell:=\ell+1$

\noindent\textbf{variable to check} : For each variable node $v = 1,2,\ldots,N$, let $\partial_v$ be the set of adjacent check nodes of $v$.
The message $p_{vc}^{(\ell)} \in \GF(2^m)$ sent from $v$ to $c$ is computed as follows
\begin{center}
$p_{vc}^{(\ell)}(x) = \xi p_{v}^{(0)}(x) \prod_{c' \in \partial_v\setminus\{c\}}  p_{c'v}^{(\ell)}(x)$ for $x \in \GF(2^m)$,
\end{center}
where $\xi$ is the normalized constant so that 
$\sum_{x \in \GF(2^m) } p_{vc}^{(\ell)}(x) = 1$.

\noindent\textbf{tentative decision} : For $v=1,2,\ldots,N$, the tentative decision of the $v$-th symbol is given by 
\begin{align*}
 \hat{x}_{v}^{(\ell)} = \mathop {\arg \max }\limits_{x \in \GF(2^m )} p_v^{(0)}(x)\prod_{c \in \partial_v}  p_{cv}^{(\ell)}(x),
\end{align*}
Let $\hat{\mathbf{x}} = (\hat{x}^{(\ell)}_1,\hat{x}^{(\ell)}_2,\ldots,\hat{x}^{(\ell)}_N)$ be the estimated codeword of iteration round $\ell$.
The decoder stops when the maximum iteration $\ell_{\textrm{max}}$ is reached or
$\mathbf{H}\hat{\mathbf{x}} = \mathbf{0} \in \GF(2^m)^M$. 
Otherwise repeat the latter 3 decoding steps.
After the decoder stops, $\RN$ sends $\hat{\mathbf{x}}$ as the codeword in MAC mode.
\section{Relay Channel}
This paper deals with the \textit{time-division} half-duplex relay channel which employs the decode-and-forward protocol.
Throughout this paper, we refer to the time-division half-duplex relay channel with the decode-and-forward protocol 
as the relay channel for the sake of simplicity.

\subsection{Time-Division Half Duplex Relay Channel}
\begin{figure}[htb]
\centering
\includegraphics[scale=0.9]{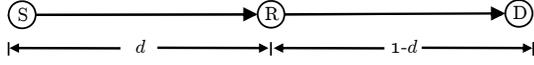}
\caption{A relay system with a source, a relay and a destination denoted by $\SN$, $\RN$ and $\DN$ respectively. The relay is on the direct line between $\SN$ and $\DN$.}
\label{model}
\end{figure}

\begin{figure}[htb]
\centering
\includegraphics[scale=0.9]{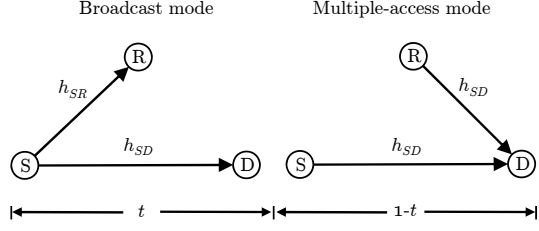}
\caption{The transmission in the time-division half-duplex relay channel.}
\label{time_division}
\end{figure}

In half-duplex relaying, a relay cannot transmit and receive simultaneously in one time slot so the concept of time-division is introduced \cite{relay_chan5,td}.
In our model, we assume that the relay $\mathrm{(R)}$ places on the direct line 
between a source $\mathrm{(S)}$ and a destination $\mathrm{(D)}$ as shown in Fig. \ref{model}.
The distance between $\SN$ and $\DN$ is normalized to 1 and let $d$ denote the distance 
between $\SN$ and $\RN$.
The transmission in relay channel takes place over two time slots of normalized duration.
In BC mode, $\SN$ sends information to $\RN$ and $\DN$ within the first time slot of duration $t$.
We refer a parameter $t$ as time-sharing factor.
In the second time slot of duration $t' = 1 - t$, $\RN$ and $\SN$ send information to $\DN$
and we refer to the operations in this time slot as MAC mode.
Figure \ref{time_division} illustrates the transmission in the time-division half-duplex relay channel.
Parameters $h_{\SD}$, $h_{\SR}$ and $h_{\RD}$ generally represent the channel effects 
such as fading, shadowing and path loss between two terminals.
In this paper, we consider only the large scale path loss.
Under this circumstance, $|h_{\SD}|^{2}=1$, $|h_{\SR}|^{2}=\frac{1}{d^{\alpha}}$ 
and $|h_{\RD}|^2=\frac{1}{(1-d)^{\alpha}}$ 
where $\alpha$ is the path loss exponent.

For a fair comparison with direct transmission (the communication that consists of $\SN$ and $\DN$),
the transmitted powers of $\SN$ and $\RN$ must obey an average global power constraint given as \cite{LDPC_relay1}
\begin{equation}
\label{power_con}
\Theta : tP_{\SN,\BC} + t'(P_{\SN,\MAC}+P_{\RN,\MAC}) \leq P,
\end{equation}
where $P_{\SN,\BC}$ is the average transmitted power of $\SN$ in BC mode, 
$P_{\SN,\MAC}$ and $P_{\RN,\MAC}$ are the average transmitted powers of $\SN$ and $\RN$ in MAC mode respectively. 
We define $P$ as the total transmission power of the relay system.
The alphabet in subscript is used to describe the position of variables under consideration.

For $i$-th transmission,
let $s_{\SN,\BC}^{(i)}$ be the modulated binary signals transmitted from $\SN$ in $\BC$ mode,
let $s_{\SN,\MAC}^{(i)}$ be the modulated binary signals transmitted from $\SN$ in $\MAC$ mode
and let $s_{\RN,\MAC}^{(i)}$ be the modulated binary signals transmitted from $\RN$ in $\MAC$ mode.
In the similar way, $y_{\DN,\BC}^{(i)}$, $y_{\DN,\MAC}^{(i)}$ and $y_{\RN,\BC}^{(i)}$ are defined as the received signals 
at given positions and modes.
From the notations above, the received signals at $\RN$ and $\DN$ in BC mode can be written as
\begin{equation}
\label{eq2}
y_{\DN,\BC}^{(i)} = h_{\SD}s_{\SN,\BC}^{(i)} + n_{\DN,\BC}^{(i)},
\end{equation}
\begin{equation}
\label{eq3}
y_{\RN,\BC}^{(i)} = h_{\SR}s_{\SN,\BC}^{(i)} + n_{\RN,\BC}^{(i)},
\end{equation}
In MAC mode, the received signal at $\DN$ is given as
\begin{equation}
\label{eq4}
y_{\DN,\MAC}^{(i)} = h_{\SD}s_{\SN,\MAC}^{(i)} + h_{\RD}s_{\RN,\MAC}^{(i)} + n_{\DN,\MAC}^{(i)}
\end{equation}
where $n_{\DN,\BC}^{(i)}$, $n_{\RN,\MAC}^{(i)}$ and $n_{\DN,\MAC}^{(i)}$ are AWGN with zero mean and unit variance at given positions and modes.
In this paper, 
%we normalize the noise power to unity so that $P$ can be used to represent signal to noise ratio (SNR).
the relationship between the total transmission power and the signal to noise ratio (SNR) of the relay system is defined as follows \cite{LDPC_relay1}.
\begin{equation}
\label{SNR_def}
\begin{array}{cl}
   {\textrm{SNR}} & { = tP_{\SN,\BC} + t'(P_{\SN,\MAC}+P_{\RN,\MAC})}  \\
   {} & { = 2R_\mathrm{r}\mathrm{{E_{b}/N_{0}}} },  \\
\end{array}
\end{equation}
where $R_\mathrm{r}$ is the coding rate for the relay channel and $\mathrm{{E_{b}/N_{0}}}$ is the the energy per bit to noise power spectral density ratio.
Since the power of noise is normalized to unity, the SNR at $\SN$ in both BC and MAC modes can be expressed as follows
\begin{align}
\label{eq4}
&\textrm{SNR}_{\DN,\BC} = |h_{\SD}|^2 P_{\SN,\BC},\\
\label{eq4}
&\textrm{SNR}_{\DN,\MAC} = \left( h_{\SD}\sqrt{P_{\SN,\MAC}} + h_{\RD}\sqrt{P_{\RN,\MAC}} \right)^2. 
\end{align}

\subsection{Decode-and-Forward Protocol}
We summarize the common parameters before describing the decode-and-forward protocol.
Let $K$ denote the information length in symbol. 
Let $N$ denote the length of the code used at $\SN$ for BC mode. 
Let $N'$ denote the length of the codes used for MAC mode.
The relationship between the time-sharing factor and the lengths of codes used in BC and MAC modes 
can be expressed as $\frac{N}{N'} = \frac{t}{t'}$.
The overall coding rate of the relay channel is given by $R_{\mathrm{r}}=\frac{K}{N + N'}$ 
%We divide $\mathscr{N}$ coded symbols into two parts which are $t\mathscr{N}$ and $t'\mathscr{N}$ 
%where $t\mathscr{N} + t'\mathscr{N} = \mathscr{N}$.
%The remaining $N_2$ symbols will be subsequently transmitted in MAC mode 

In BC mode, $\SN$ firstly encodes the $K$  information symbols 
with a code of length $N$ and of rate $R_{\SN,\BC}=\frac{K}{N}$. 
The $N$ coded symbols are transmitted to both $\RN$ and $\DN$.
%$\RN$ enables  to decode the $K$ information symbols from the received signals since $|h_{\SR}|^{2}>|h_{\SD}|^{2}$.
$\RN$ decodes its received signals and the estimation obtained from decoder will be further used in MAC mode.
Meanwhile, $\DN$ just stores the received signals.
In MAC mode, $\RN$ transmits another set of $N'$ symbols to $\DN$.
%has an option to re-encode and then
At the same time, $\SN$ also simultaneously sends $N'$ symbols to $\DN$. 
Finally, $\DN$ can decode the received information transmitted in BC mode 
at lower coding rate $R_{\DN,\MAC}=\frac{K}{N+N'}=R_{\mathrm{r}}$ since it receives the additional information transmitted in the MAC mode. 
This implies that the computational complexity of decoder at $\DN$ is much higher than that of decoder at $\RN$ 
since the number of parity symbols at $\DN$ is more than the number of parity symbols at $\RN$.
This is more crucial when the overall relay system operates at low coding rate.
In this paper, we will develop coded relay system which $\RN$ and $\DN$ can decode their received signals with almost the same computational complexity.

\subsection{Achievable Rate}
\begin{figure}[htb]
\centering
\includegraphics[scale=0.65]{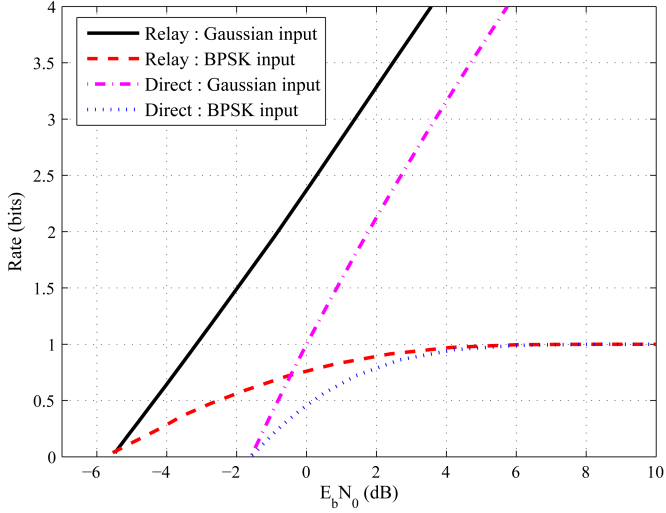}
\caption{The achievable rate of a relay channel for $d=0.5$ and $\alpha=2$ obtained from optimal relay parameters and the capacity of direct transmission.}
\label{acheivable_rate}
\end{figure}

For a Gaussian relay channel, a general time-division half-duplex relay channel with decode-and-forward protocol can achieve the following rate \cite{LDPC_relay1}.
\begin{equation}
\label{acheivable_rate_eq}
\mathcal{R} = \mathop {\sup }\limits_{\Theta, 0\le t,r \le 1} \min \left\{ {t\mathcal{C}\left( w \right) + t'\mathcal{C}\left( x \right),t\mathcal{C}\left( y \right) + t'\mathcal{C}\left( z \right)} \right\},
\end{equation}
where
\begin{align*}
   w =&  {\left| {h_{\SR} } \right|^2 P_{\SN,\BC}},  \\
   x =&  {\left( {1 - r^2 } \right)\left| {h_{\SD}} \right|^2 P_{\SN,\MAC} },  \\
   y =&  {\left| {h_{\SD} } \right|^2 P_{\SN,\BC} },  \\
   z =&  {\left| {h_{\SD} } \right|^2 P_{\SN,\MAC}  + \left| {h_{\RD} } \right|^2 P_{\RN,\MAC} }  \\
    &  { + 2r\sqrt {\left| {h_{\SD} } \right|^2 \left| {h_{\RD} } \right|^2 P_{\SN,\MAC} P_{\RN,\MAC}}},  \\
   C(\textrm{SNR}) & = \mathrm{log}_2(1+\textrm{SNR}),
\end{align*}
and $\Theta$ is defined in (\ref{power_con}). The parameter $r$ represents the correlation between $s_{\SN,\MAC}$ and $s_{\RN,\MAC}$ in MAC mode.
The function $\mathcal{C}(\cdot)$ is known as the Shannon formula used to calculate capacity of Gaussian link. 
In order to achieve $\mathcal{R}$ given in (\ref{acheivable_rate_eq}), a joint optimization is needed over the correlation in MAC mode, time-sharing factor and power allocation.
Figure \ref{acheivable_rate} compares the capacity of the direct transmission with the achievable rate of a relay channel with $d=0.5$ and $\alpha=2$.
In this figure, we plotted the coding rate  versus $\mathrm{{E_{b}/N_{0}}}$ which is the normalized $\textrm{SNR}$.
Details of these calculations are fully presented in \cite{LDPC_relay1,acheiveR1}.
This figure clearly shows that the achievable rate of the relay channel is much higher than the capacity of direct transmission especially for low $\textrm{SNR}$.
Therefore, we focus our attention on designing  low-rate NBLDPC codes for the relay channel.
%\textcolor{red}{This implies the importance of coding the relay channel of low capacity. 
%}
%\textcolor{red}{[Puripong, how about focusing only on lower-rate schemes?]}

%However, designing of a transceiver for the relay channel becomes more difficult and complicated under optimal relay parameters.
%Therefore, some practical constraints such as silent source in MAC mode, equal power allocation or equal time-sharing can be considered.
%Theses constraints greatly simplify the scheduling, synchronization as well as the design of coding strategy of the relay channel.

\section{Low-Rate Relay Coding Strategy}

In this section, we propose a novel low-rate NBLDPC coding strategy for the relay channel.
The proposed coding strategy is analogous to constructing $C_T$ of the multiplicatively repeated NBLDPC code.
For this strategy, the relationship between time sharing factor $t$ and repetition parameter $T$ are $t=\frac{1}{T}$ 
and hence $t'=1-\frac{1}{T}$.
For $x_v \in \GF(2^m) $, $v=1,2,\ldots,N$ and $i=1,2,\ldots,m$, $\RN$,
let $Y_{\RN,\BC}^{((v-1)m+i)}$ be the random variable 
with the corresponding received signal $y^{((v-1)m+i)}_{\RN,\BC}$.
We also define the random variables $Y_{\DN,\BC}^{((v-1)m+i)}$ and $Y_{\DN,\MAC}^{((v-1)m+i)}$ 
and these variables are interpreted similarly to the definition of $Y^{((v-1)m+i)}_{\RN,\BC}$.
By using the decode-and-forward protocol, 
we describe the low-rate NBLDPC coding strategy for the relay channel as follows.

%%%%%%%%%%%%%%%%%%%%%%%%
\textit{1) Encoding in BC mode:}
By using an NBLDPC code of rate $R_{\SN,\BC}=\frac{K}{N}$ as $C_1$, 
$\SN$ encodes $K$ information symbols into codeword $\mathbf{x} = \left( x_1,x_2,\ldots,x_{N} \right)\in \GF(2^m)^N$.
$\SN$ sends $\mathbf{x}$ to both $\RN$ and $\DN$.
For this strategy, $\SN$ also stores the codeword $\mathbf{x} \in C_1$ for the next transmission in MAC mode. 
Subsequently $\mathbf{x}$ is mapped to a binary sequence of length $mN$ bits.  Then the BPSK modulated signals of the binary sequence are sent to $\RN$ and $\DN$ at the same time.

%%%%%%%%%%%%%%%%%%%%%%%%
\textit{2) Decoding in BC mode:}
$\RN$ decodes its received signals by using the BP algorithm (described in section II-D).
%For $x_v \in \GF(2^m) $, $v=1,2,\ldots,N$ and $i=1,2,\ldots,m$, 
$\RN$ first calculates $p_{v}^{(0)}(x)$ from $y^{((v-1)m+1)}_{\RN,\BC},\dotsc,y^{(vm)}_{\RN,\BC}$ as follows. 
\begin{align*}
p_{v}^{(0)}(x)&=\prod_{i=1}^m\Pr( X_{v,i} = x_{i} \mid Y_{\RN,\BC}^{((v-1)m+i)}=y^{((v-1)m+i)}_{\RN,\BC}) \\ 
&=\prod_{i=1}^m \frac{1}{\sqrt{2\pi}}\mathrm{exp}(-\frac{1}{2} ( y^{((v-1)m+i)}_{\RN,\BC} - h_{\SR}s_i )^2),
\end{align*}
where $(x_1,\dotsc,x_m)\in\GF(2)^m$ is the binary representation of $x\in\GF(2^m)$,
and $s_i$ is the modulates binary signal corresponding to $x_i$. 
Then the decoder at $\RN$ is initialized $p_{v}^{(0)}(x)$ as the input of the BP algorithm.
 
After BP decoding, $\RN$ produces an estimated codeword $\hat{\mathbf{x}}_{\RN}\in\GF(2^m)^N$.
Meanwhile, $\DN$ only stores its received signals for the future decoding at the end of MAC mode.

%%%%%%%%%%%%%%%%%%%%%%%%
\textit{3) Encoding in MAC mode:}
The encoding processes for both $\SN$ and $\RN$ in this mode are analogous to constructing the code $C_T$ from multiplicative repetition.
By using $\mathbf{x} = \left( x_1,x_2,\ldots,x_{N} \right) \in C_1$ stored in BC mode,
$\SN$ produces a codeword of code $C_T$ by multiplicative repetition as described in Section II-B.

$\SN$ then sends only the multiplicative symbol 
$\mathbf{x_{\SN}} = \left( x_{N+1},x_{N+2},\ldots,x_{TN} \right)$ 
of length $N'=(T-1)N$ symbols to $\DN$, where 
$x_{(t-1)N+v} = r_{(t-1)N+v}x_{v},  \text{ for } v=1,\ldots,N, t=2,\dotsc,T$, 
and  $r_{N+1},\dotsc,r_{TN}$ are randomly chosen from $\GF(2^m)\setminus \{0\}$.

In the same way, as $\SN$ produced $\mathbf{x_{\SN}}$ from $\mathbf{x}$,
$\RN$ produces $\mathbf{x_{\RN}}\in\GF(2^m)^{(T-1)N}$ from $\hat{\mathbf{x}}_{\RN}\in\GF(2^m)^{N}$.
We note that $\SN$ and $\RN$ send the same coded symbols $\mathbf{x_{\SN}}=\mathbf{x_{\RN}}$  if $\RN$, in BC mode, did successfully decode 
the codeword, i.e., $\mathbf{x}=\hat{\mathbf{x}}$.
This means the additional parity symbols sent from $\SN$ and $\RN$ in this mode are fully correlated $(r=1)$.
Thus, the additional $(T-1)N$ symbols are sent to $\DN$.

%%%%%%%%%%%%%%%%%%%%%%%%
\textit{4) Decoding in MAC mode:}
At this step, $\DN$ has two received signals which are 
$y_{\DN,\BC}^{(1)},\dotsc,y_{\DN,\BC}^{(mN)}$ 
and 
$y_{\DN,\MAC}^{(mN+1)},\dotsc,y_{\DN,\MAC}^{(mTN)}$ 
which correspond to the overall $TN$ coded symbols or equivalently $mTN$ coded bits.
Therefore, the overall coding rate of the relay channel is $R_\mathrm{r} = \frac{K}{TN} = \frac{R_{\SN,\BC}}{T}$.
The received signals $y_{\DN,\BC}^{(1)},\dotsc,y_{\DN,\BC}^{(mN)}$ correspond to the coded symbol of length $N$ transmitted in BC mode.
whereas the received signals $y_{\DN,\MAC}^{(mN+1)},\dotsc,y_{\DN,\MAC}^{(mTN)}$ 
correspond to the coded symbol of length $(T-1)N$ transmitted in MAC mode.
Let $X_{v}$ be the random variables with realization $x_{v}$  where $X_{v}$ represents the overall coded symbol transmitted from both BC and MAC modes and $v=1,2,\ldots,TN$.
Let $Y_{v}$ be the random variables with realization $y_{v}$ 
which is the received value from both BC and MAC modes.
The probability of transmitted symbol $\Pr(X_{v})$ is assumed to be uniform.
From $y_{\DN,\BC}^{(1)},\dotsc,y_{\DN,\BC}^{(mN)}$ and $y_{\DN,\MAC}^{(mN+1)},\dotsc,y_{\DN,\MAC}^{(mNT)}$,
the decoder at $\DN$ first calculates $p_{v}^{(0)}(x)$ for $v=1,\dotsc,N$ as follows.
\begin{align*}
&p_{v}^{(0)}(x)\\
&=\xi\Pr (X_{v}  = x\mid Y_{\DN,\BC}^{((v-1)m+i)}=y_{\DN,\BC}^{((v-1)m+i)} \\
&\hspace{3cm}\text{ for } i=1,\dotsc,m)  \\
   \cdot&\phantom{\xi}\prod_{t=1}^{T-1}\Pr (X_{tN+v}  = r_{tN+v}x \mid  \\
&\hspace{1cm}Y_{\DN,\MAC}^{((tN+v)m+i)}=y_{\DN,\MAC}^{((tN+v)m+i)}\text{ for } i=1,\dotsc,m)  \\
 &=\xi\prod_{i=1}^m\frac{1}{{\sqrt {2\pi } }}\exp ( - \frac{1}{2}(y_{\DN,\BC}^{((v - 1)m + i)}  - h_{\SN\DN}s_{i} ))^2 \\
 & \prod_{t=1}^{T-1}\prod_{i=1}^m \frac{1}{{\sqrt {2\pi } }}\exp ( - \frac{1}{2}(y_{\DN,\MAC}^{(tN+(v - 1)m + i)}-(h_{\SD}+h_{\RD})s_{t,v,i})^2),
\end{align*}
where $s_{t,v,i}$ is the modulated binary signal corresponding to  the $i$-th bit of the binary representation of $r_{tN+v}x_v\in\GF(2^m)$,
and $\xi$ is  the normalized constant so that $\sum_{x \in \GF(2^m) } p_{v}^{(0)}(x) = 1$.
Then the decoder at $\DN$ is initialized $p_{v}^{(0)}(x_v)$ as the input of the BP algorithm.
After BP decoding, $\DN$ produces an estimated codeword $\hat{\mathbf{x}}_{\DN}\in\GF(2^m)^N$.

The initial calculation of $p_{v}^{(0)}(x)$ can be viewed as being calculated by
combining two received signals $y_{\DN,\BC}^{(i)}$ and $y_{\DN,\MAC}^{(i)}$.
After the two received signals are combined, the decoder performs the BP algorithm on the Tanner graph of $C_1$ with $N$ variable nodes.
%Finally, $\DN$ produces the estimated codeword $\hat{\mathbf{x}}_{\DN} = \left( \hat{x}_1,\hat{x}_2,\ldots,\hat{x}_{N} \right)$ as the output.
%\begin{center}
%\line(1,0){250}
%\end{center}

\begin{figure}[htb]
\centering
\includegraphics[scale=1.35]{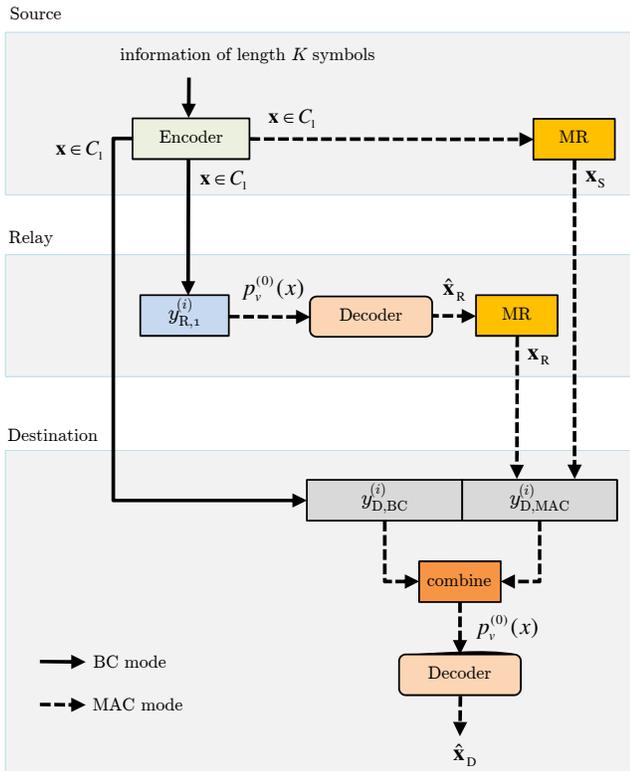} 
\caption{Block diagram of the proposed low-rate NBLDPC coding strategy for the relay channel. 
In this figure, the block labeled with ``MR" represents multiplicative repetition as described in Section IV.
The block labeled with ``combine" represents the mathematical operations according to the initialization step of decoding at $\DN$. 
The Tanner graphs at the decoders at $\RN$ and $\DN$ are the same.}
\label{nb_coding_relay}
\end{figure}

In summary, 
the proposed low-rate NBLDPC coding strategy for the relay channel is depicted in Fig. \ref{nb_coding_relay}.
We claim that the proposed coding strategy is very simple due to the following two reasons.
1) For encoding in MAC mode, the encoder at $\RN$ uses only multiplicative repetition which is much simpler than encoders of BLDPC codes.
The multiplicative repetition requires only $(T-1)N$ multiplications over $\GF(2^m)$ for encoding.
2) The Tanner graphs of decoders at both $\RN$ and $\DN$ are the same, i.e., the Tanner graph of $C_1$. 
The difference of decoding at $\RN$ and $\DN$ is only the initialization. 
And the computational complexity of the decoders are almost the same
even if the coding rate at $\DN$ was lower than that of $\RN$. 

\section{Numerical Results}
In this section, we demonstrate the decoding performance of NBLDPC coded relay system.
For all results, 
the NBLDPC codes defined over $\GF(2^8)$ with regular degree profile $(\dv=2,\dc)$ are employed.
We choose $m=8$ for its good performance and also the computer-friendliness of byte.
The FFT-based BP algorithm \cite{nb2} 
with the maximum iteration $\ell_{\textrm{max}}=500$ is used for decoding.
The BPSK modulation was used for all the links in the relay channel.
We set the attenuation exponent to $\alpha=2$ and the source-relay distance to $d=0.5$.
We denote the bit error rate and frame error rate as BER and FER respectively.

First, we present the BER performances of the proposed low-rate coding strategy for the relay channel.
The power allocation for $\SN$ and $\RN$ are static and  fixed as follows:
$P_{\SN,\BC}= \frac{P}{2}$, $P_{\SN,\MAC}=\frac{P}{4}$ and $P_{\RN,\MAC}=\frac{P}{4}$.
A $(\dv=2,\dc=3)$-regular NBLDPC codes of rate $R=\frac{1}{3}$ over $\GF(2^8)$ are chosen as the mother code $C_1$.
We choose the repetition parameter $T=2$ for encoding in MAC mode.
This corresponds to the {\it equal time sharing constraint} since $t=\frac{1}{T} = \frac{1}{2}$.
By using this strategy, the overall coding rate is given by $R_{\mathrm{r}} = \frac{R_{\SN,\BC}}{T} = \frac{R}{T} = \frac{1}{6}$.

\begin{figure}[htb]
\centering
\includegraphics[scale=0.63]{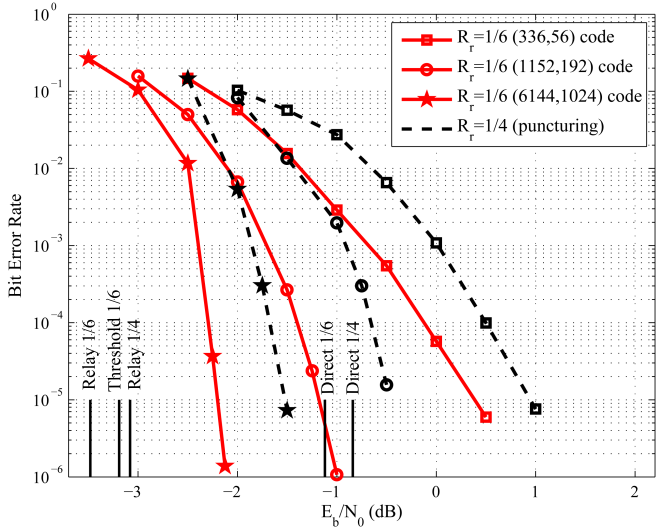} 
\caption{The BER performances of NBLDPC coded relay systems for low-rate case. 
The parameters in parentheses represent the overall transmitted bits and the information bits respectively. 
In this figure, the word ``Relay" represents achievable rate of relay channel 
and the word ``Direct" represents the Shannon's capacity of direct transmission.
The threshold value for $R_{\mathrm{r}} = \frac{1}{6}$ is represented by ``Threshold".
The black curves labelled with ``puncturing" represent the rate-compatible relay systems
with $R_{\mathrm{r}}=\frac{1}{4}$ obtained by puncturing the codes of rate $R_{\mathrm{r}}=\frac{1}{6}$.}
\label{BER_results}
\end{figure}

Figure \ref{BER_results} shows the BER performances of the proposed three low-rate NBLDPC codes for the relay channel. 
The three NBLDPC codes have information length $k\in\{56,192,1024\}$ bits, respectively.
The threshold value for the proposed coding strategy with $R_{\mathrm{r}}=\frac{1}{6}$ is calculated by the Monte Carlo density evolution.
The Monte Carlo density evolution originally developed in \cite{MC} is a method to find the threshold of NBLDPC codes in the limit of very large codeword length.
It can be seen that the proposed codes theoretically perform very close to the achievable rate. The gap of threshold from the achievable rate is only $0.3$ dB.
We can see from the figure that BER curve improves by  increasing of information length (also the codeword length).
% These results confirm that the threshold value of $R_{\mathrm{r}}=\frac{1}{6}$ is validity since the BER curve of 
% $R_{\mathrm{r}}=\frac{1}{6}$ tends to move toward to the threshold when the codeword length is increased.
At BER of $10^{-5}$ and $R_\mathrm{r} = \frac{1}{6}$, it can be seen from the figure that the proposed NBLDPC code of information length $k=1024$ bits  
performs within $1.2$ dB from the achievable rate.
Moreover, the coding gain is about $1.1$ dB beyond  the capacity of direct transmission.

%The performance of a rate-compatible relay system is also considered in this paper.
%We note again that the rate-compatible relay system can support different coding rates by adopting only one channel code.
We explain how the proposed strategy can be efficiently used for the rate-compatible relay system 
by investigating the performances of relay systems with $R_\mathrm{r}=\frac{1}{4}$ 
obtained from puncturing the codes used in relay systems with $R=\frac{1}{6}$.
%We consider the case that we adopt only $R=\frac{1}{3}$ NBLDPC codes as the mother codes.
By using the {\itshape recoverable step} discussed in Section II-C, 
we puncture the (2,3)-regular mother NBLDPC code of rate $R=\frac{1}{3}$  to obtain an NBLDPC code of rate $R_{\SN,\BC}=\frac{1}{2}$.
For example, we obtain a $(N=48,K=24)$ NBLDPC code by puncturing $(N=72, K=24)$ NBLDPC code.
This means that we cancel $N_p=24$ parity symbols of $(N=72,K=24)$ NBLDPC code.
With repetition parameter $T=2$ at $\RN$, 
the overall coding rate is $R_{\mathrm{r}} = \frac{R_{\SN,\BC}}{T} = \frac{1}{4}$.
The BER performances of the proposed strategy with punctured NBLDPC codes (dashed curves) are shown in Fig. \ref{BER_results}.
Although the codes are punctured but
the NBLDPC coded relay system of rate $R=\frac{1}{4}$ still obtain a coding gain around $0.75$ dB 
beyond the capacity of direct transmission.
%Therefore, the proposed strategy is also applicable for rate-compatible relay system.

\begin{figure}[htb]
\centering
\includegraphics[scale=0.63]{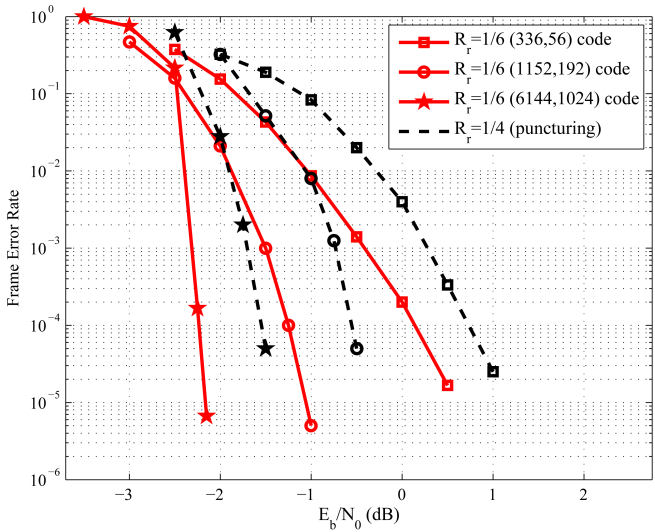} 
\caption{The FER performance of NBLDPC coded relay systems for low-rate case.}
%The parameters in parentheses represent the overall transmitted bits and the information bits respectively.
%Like Fig. \ref{BER_results}, the black curves labelled with ``puncturing" represent the rate-compatible relay systems
%with $R_{\mathrm{r}}=\frac{1}{4}$ obtained from $R_{\mathrm{r}}=\frac{1}{6}$.
\label{FER_results}
\end{figure}

As mentioned in \cite{LDPC_relay1}, the optimized BLDPC codes need the concatenation with outer BCH or RS codes to guarantee the good FER performance.
The advantage of using $(2,\dc)$-regular NBLDPC codes over the direct transmission is that 
the regular NBLDPC codes with $\dv=2$ have a good FER performance at both waterfall and error region \cite{FER_good1,FER_good2}.
Figure \ref{FER_results} shows the FER performances of NBLDPC codes for the relay channel. 
The proposed low-rate codes exhibit excellent FER performances at both waterfall and error region without concatenating BCH or RS codes.
The punctured NBLDPC coded relay systems with $R=\frac{1}{4}$ (dashed curve) exhibit
very good FER performances.

%%In this section, we have demonstrated that our proposed NBLDPC coding strategies run the relay systems effectively
%%at both short and moderate codeword lengths.
%%Therefore, NBLDPC codes are the promising candidate for using as the channel code of relay systems.

%\section{Conclusion}
%\textcolor{blue}{
%% identify idea
%In this paper, we have presented the use of the NBLDPC codes for the relay system. 
%The relay systems are coded with the $(\dv=2,\dc)$-regular NBLDPC codes defined over $\GF(2^8)$.
%% make ur contribution explicit
%We propose two simple NBLDPC coding strategies for both low and high rate relay channels.
%% restate our main finding
%Our results clearly show that the NBLDPC codes are the capacity-approaching code for decode-and-forward half-duplex relay channel.
%The NBLDPC codes also exhibit the better performance comparing with BLDPC codes even with the shorter codeword length.
%% benefit
%Moreover, very good decoding performances are obtained with simple decoder.
%We propose the decoding methods that the relay and the destination can decode at almost the same complexity 
%by sharing the same structure of decoder.}

\section{Conclusion}
We propose in this paper the low-rate NBLDPC coding strategy for the decode-and-forward half-duplex relay channels.
The advantages of the proposed strategy can be listed as follows:
1) The waterfall performance within $1.5$ dB from the achievable rate can be obtained from the moderate length NBLDPC codes.
The significant coding gain is also achieved over the Shannon's limit of the direct transmission.
2) We do not need the optimization process for optimizing the degree profile of LDPC codes. 
We just employ the $(2,\dc)$-regular NBLDPC defined over $\GF(2^8)$ as the channel code.
3) The FER performances obtained from the $(2,\dc)$-regular NBLDPC codes are excellent even the outer BCH or RS codes are not employed to lower the error floors.
4) The relay and the destination can decode with almost the same computational complexity 
since both the relay and the destination use the same Tanner graph for decoding.
5) The encoding processes for both the source and the relay in MAC mode are very simple 
since we use the multiplicative repetition which requires only $(T-1)N$ multiplications over $\GF(2^m)$.
6) The proposed strategy is applicable for rate-compatible relay systems.
By adapting the repetition parameter $T$ at the relay or puncturing the codes at the source, easy HARQ is possible for very noisy relay channels.

% use section* for acknowledgement
\section*{Acknowledgment}
The authors would like to thank the Department of Electrical Engineering, Khon Kaen University.
This work is financially supported by the Telecommunications Research Industrial and Development Institute (TRIDI), with National Telecommunications Commission (NTC), Grant No.MS/009/2552. 
The authors also would like to acknowledge the discussions and guidance of M. Wu.

%%% --- references section --- %%%

\bibliographystyle{IEEEtran}
`\bibliography{IEEEabrv,my_references}

% Generated by IEEEtran.bst, version: 1.13 (2008/09/30)
\begin{thebibliography}{10}
\providecommand{\url}[1]{#1}
\csname url@samestyle\endcsname
\providecommand{\newblock}{\relax}
\providecommand{\bibinfo}[2]{#2}
\providecommand{\BIBentrySTDinterwordspacing}{\spaceskip=0pt\relax}
\providecommand{\BIBentryALTinterwordstretchfactor}{4}
\providecommand{\BIBentryALTinterwordspacing}{\spaceskip=\fontdimen2\font plus
\BIBentryALTinterwordstretchfactor\fontdimen3\font minus
  \fontdimen4\font\relax}
\providecommand{\BIBforeignlanguage}[2]{{%
\expandafter\ifx\csname l@#1\endcsname\relax
\typeout{** WARNING: IEEEtran.bst: No hyphenation pattern has been}%
\typeout{** loaded for the language `#1'. Using the pattern for}%
\typeout{** the default language instead.}%
\else
\language=\csname l@#1\endcsname
\fi
#2}}
\providecommand{\BIBdecl}{\relax}
\BIBdecl

\bibitem{coop1}
J.~Laneman, D.~Tse, and G.~Wornell, ``Cooperative diversity in wireless
  networks: Efficient protocols and outage behavior,'' \emph{{IEEE} Trans. Inf.
  Theory}, vol.~50, no.~12, pp. 3062 -- 3080, Dec. 2004.

\bibitem{coop2}
G.~Kramer, M.~Gastpar, and P.~Gupta, ``Cooperative strategies and capacity
  theorems for relay networks,'' \emph{{IEEE} Trans. Inf. Theory}, vol.~51,
  no.~9, pp. 3037 -- 3063, Sep. 2005.

\bibitem{coop3}
A.~Sendonaris, E.~Erkip, and B.~Aazhang, ``User cooperation diversity. {P}art
  {I} and {II}.'' \emph{{IEEE} Trans. Commun.}, vol.~51, no.~11, pp. 1927 --
  1948, Nov. 2003.

\bibitem{diversity}
D.~Tse and P.~Viswanath, \emph{Fundamentals of Wireless Communication}.\hskip
  1em plus 0.5em minus 0.4em\relax Cambridge University Press, Mar. 2005.

\bibitem{mimo1}
A.~Paulraj, D.~Gore, R.~Nabar, and H.~Bolcskei, ``An overview of {MIMO}
  communications - a key to gigabit wireless,'' \emph{Proceedings of the IEEE},
  vol.~92, no.~2, pp. 198 -- 218, Feb. 2004.

\bibitem{mimo2}
Q.~Li, G.~Li, W.~Lee, M.~il~Lee, D.~Mazzarese, B.~Clerckx, and Z.~Li, ``{MIMO}
  techniques in {WiMAX} and {LTE}: a feature overview,'' \emph{{IEEE} Commun.
  Mag.}, vol.~48, no.~5, pp. 86 --92, May 2010.

\bibitem{mimo_coop1}
C.-X. Wang, X.~Hong, X.~Ge, X.~Cheng, G.~Zhang, and J.~Thompson, ``Cooperative
  {MIMO} channel models: A survey,'' \emph{{IEEE} Commun. Mag.}, vol.~48,
  no.~2, pp. 80 --87, Feb. 2010.

\bibitem{mimo_coop2}
Y.~Fan and J.~Thompson, ``{MIMO} configurations for relay channels: Theory and
  practice,'' \emph{{IEEE} Trans. Wireless Commun.}, vol.~6, no.~5, pp. 1774
  --1786, May 2007.

\bibitem{relay_chan1}
E.~C. van~der Meulen, ``Three-terminal communication channels,'' \emph{Advanced
  Applied Probability}, vol.~3, pp. 120--154, 1971.

\bibitem{relay_chan2}
T.~Cover and A.~Gamal, ``Capacity theorems for the relay channel,''
  \emph{{IEEE} Trans. Inf. Theory}, vol.~25, no.~5, pp. 572 -- 584, Sep. 1979.

\bibitem{relay_chan3}
Y.~Ding, J.-K. Zhang, and K.~Wong, ``Ergodic channel capacities for the
  amplify-and-forward half-duplex cooperative systems,'' \emph{{IEEE} Trans.
  Inf. Theory}, vol.~55, no.~2, pp. 713 --730, Feb. 2009.

\bibitem{relay_chan4}
S.~Salehkalaibar, L.~Ghabeli, and M.~Aref, ``An achievable rate for relay
  networks based on compress-and-forward strategy,'' \emph{{IEEE} Commun.
  Lett.}, vol.~14, no.~4, pp. 279 --281, Apr. 2010.

\bibitem{relay_chan5}
A.~Host-Madsen and J.~Zhang, ``Capacity bounds and power allocation for
  wireless relay channels,'' \emph{{IEEE} Trans. Inf. Theory}, vol.~51, no.~6,
  pp. 2020 --2040, Jun. 2005.

\bibitem{relay_chan6}
M.~Wu, P.~Weitkemper, D.~Wubben, and K.-D. Kammeyer, ``Comparison of
  distributed ldpc coding schemes for decode-and-forward relay channels,'' in
  \emph{International ITG Workshop on Smart Antennas (WSA)}, Feb. 2010, pp. 127
  --134.

\bibitem{code_relay1}
Z.~Zhang and T.~Duman, ``Capacity-approaching turbo coding for half-duplex
  relaying,'' \emph{{IEEE} Trans. Commun.}, vol.~55, no.~10, pp. 1895 --1906,
  Oct. 2007.

\bibitem{code_relay2}
C.~Li, G.~Yue, X.~Wang, and M.~Khojastepour, ``{LDPC} code design for
  half-duplex cooperative relay,'' \emph{{IEEE} Trans. Wireless Commun.},
  vol.~7, no.~11, pp. 4558 --4567, Nov. 2008.

\bibitem{LDPC_relay1}
A.~Chakrabarti, A.~D. Baynast, A.~Sabharwal, and B.~Aazhang, ``Low density
  parity check codes for the relay channel,'' \emph{{IEEE} J. Sel. Areas
  Commun.}, vol.~25, no.~2, pp. 280 --291, Feb. 2007.

\bibitem{LDPC_relay2}
P.~Razaghi and W.~Yu, ``Bilayer low-density parity-check codes for
  decode-and-forward in relay channels,'' \emph{{IEEE} Trans. Inf. Theory},
  vol.~53, no.~10, pp. 3723 --3739, Oct. 2007.

\bibitem{LDPC_relay3}
J.~Cances and V.~Meghdadi, ``Optimized low density parity check codes designs
  for half duplex relay channels,'' \emph{{IEEE} Trans. Wireless Commun.},
  vol.~8, no.~7, pp. 3390 --3395, Jul. 2009.

\bibitem{nb1}
M.~Davey and D.~MacKay, ``Low-density parity check codes over {GF($q$)},''
  \emph{{IEEE} Commun. Lett.}, vol.~2, no.~6, pp. 165 --167, Jun. 1998.

\bibitem{nb2}
D.~Declercq and M.~Fossorier, ``Decoding algorithms for nonbinary {LDPC} codes
  over {GF}($q$),'' \emph{{IEEE} Trans. Commun.}, vol.~55, no.~4, pp. 633
  --643, Apr. 2007.

\bibitem{std1}
\emph{IEEE standard for local and metropolitan area networks , Part 16: Air
  interface for fixed and mobile broadband wireless access systems}, IEEE Std.
  802.16, 2009.

\bibitem{std2}
\emph{IEEE 802.11n-2009: Wireless LAN Medium Access Control (MAC) and Physical
  Layer (PHY) Specifications: Enhancements for Higher Throughput}, IEEE Std.
  802.11, 2009.

\bibitem{nb4}
C.~Poulliat, M.~Fossorier, and D.~Declercq, ``Design of regular
  (2,$d_c$)-{LDPC} codes over {GF}($q$) using their binary images,''
  \emph{{IEEE} Trans. Commun.}, vol.~56, no.~10, pp. 1626 --1635, Oct. 2008.

\bibitem{MR1}
K.~Kasai, D.Declerq, C.~Poulliat, and K.~Sakaniwa, ``Rate-compatible non-binary
  {LDPC} codes concatenated with multiplicative repetition codes,'' in
  \emph{Proc. {IEEE} ISIT 2010}, Austin, Texas, USA, Jul. 5--9, 2010, pp. 844
  --848.

\bibitem{MR2}
K.~Kasai, D.~Declercq, C.~Poulliat, and K.~Sakaniwa, ``Multiplicatively
  repeated non-binary {LDPC} codes,'' {IEEE} Trans. Inf. Theory, Sep. 2011, to
  appear.

\bibitem{macwilliams77}
F.~J. MacWilliams and N.~J.~A. Sloane, \emph{The Theory of Error-Correcting
  Codes}.\hskip 1em plus 0.5em minus 0.4em\relax Amsterdam: Elsevier, 1977.

\bibitem{mct_book}
T.~Richardson and R.~Urbanke, \emph{Modern Coding Theory}.\hskip 1em plus 0.5em
  minus 0.4em\relax Cambridge University Press, Mar. 2007.

\bibitem{punc1}
J.~Ha, J.~Kim, D.~Klinc, and S.~McLaughlin, ``Rate-compatible punctured
  low-density parity-check codes with short block lengths,'' \emph{{IEEE}
  Trans. Inf. Theory}, vol.~52, no.~2, pp. 728 --738, Feb. 2006.

\bibitem{td}
A.~de~Baynast, A.~Chakrabarti, A.~Sabharwal, and B.~Aazhang, ``A systematic
  construction of {LDPC} codes for relay channel in time-division mode,'' in
  \emph{Fortieth Asilomar Conference on Signals, Systems and Computers, 2006.
  ACSSC '06.}, Nov. 29-1, 2006, pp. 722 --726.

\bibitem{acheiveR1}
\BIBentryALTinterwordspacing
M.~A. Khojastepour, ``Distributed cooperative communication in wireless
  networks,'' Ph.D. dissertation, Univ. of Rice, Germany, 2005. [Online].
  Available: \url{http://scholarship.rice.edu/handle/1911/20023}
\BIBentrySTDinterwordspacing

\bibitem{MC}
M.~Davey, ``Error-correction using low-density parity-check codes,'' Ph.D.
  dissertation, Univ. of Cambridge, U.K., 1999.

\bibitem{FER_good1}
H.~Song, J.~Liu, and B.~Kumar, ``Large girth cycle codes for partial response
  channels,'' \emph{{IEEE} Trans. Magn.}, vol.~40, no.~4, pp. 3084 -- 3086,
  Jul. 2004.

\bibitem{FER_good2}
X.~Tao, L.~Zheng, W.~Liu, and D.~Liu, ``Recursive design of high girth (2,$k$)
  {LDPC} codes from ($k$,$k$) {LDPC} codes,'' \emph{{IEEE} Commun. Lett.},
  vol.~15, no.~1, pp. 70 --72, Jan. 2011.

\end{thebibliography}

\end{document}